\documentclass[journal]{IEEEtran}

\usepackage[compress]{cite}
\usepackage[table]{xcolor}
\usepackage{balance}
\usepackage{pifont}
\usepackage[pdftex]{graphicx}
\DeclareGraphicsExtensions{.pdf,.jpeg,.png}

\usepackage{multirow}

\usepackage[cmex10]{amsmath}
\usepackage{amssymb}
\usepackage{optidef}
\usepackage{array}
\usepackage{mdwmath}
\usepackage{mdwtab}
\usepackage{eqparbox}
\PassOptionsToPackage{hyphens}{url}\usepackage{hyperref}
\usepackage[ruled,vlined,linesnumbered]{algorithm2e}
\usepackage{lettrine}
\usepackage{caption}
\usepackage{subcaption}
\usepackage{authblk}
\newcommand{\papertitle}{Functionality-preserving Black-box Optimization of Adversarial Windows Malware}

\newcommand{\algoname}{GAMMA\xspace}

\newcommand{\vct}[1]{\ensuremath{\boldsymbol{#1}}}
\newcommand{\mat}[1]{\ensuremath{\mathbf{#1}}}

\newcommand{\myparagraph}[1]{\noindent \textbf{#1}}
\newcommand{\ie}{i.e.\xspace}
\newcommand{\eg}{e.g.\xspace}

\newcommand{\wrt}{w.r.t.\xspace}

\newcommand{\programs}{\mathcal{X}}

\newcommand{\faspace}{\mathcal{S}}
\newcommand{\query}{q}
\newcommand{\size}{\mathcal{C}}

\newcommand{\nmalwaretest}{500\xspace}
\newcommand{\vttest}{200\xspace}

\newcommand{\cmark}{\ding{51}}%

\usepackage[draft,inline,nomargin,index]{fixme}
\fxsetup{theme=color,mode=multiuser}
\FXRegisterAuthor{aa}{envaa}{\color{red}AA}
\FXRegisterAuthor{ld}{envld}{\color{blue}LD}
\FXRegisterAuthor{gl}{envgl}{\color{orange}GL}
\FXRegisterAuthor{fr}{envfr}{\color{green}FR}
\FXRegisterAuthor{bb}{envbb}{\color{purple}BB}
\fxsetup{layout={footnote}}
\fxusetheme{colorsig}

\begin{document}
\bstctlcite{IEEEexample:BSTcontrol}
    \title{\papertitle}
\author[1]{Luca Demetrio}
\author[1,2]{Battista Biggio}
\author[3]{Giovanni Lagorio}
\author[1,2]{Fabio Roli}
\author[3]{Alessandro Armando}

\affil[1]{PRALab, Department of Electrical and Electronic Engineering,
University of Cagliari, Cagliari, Italy}
\affil[2]{Pluribus One, Italy}
\affil[3]{CSecLab, University of Genova, Genova, Italy}

\markboth{IEEE Transactions on Information Forensics and Security, VOL. N, NO. M, MONTH 2020
}{Demetrio \MakeLowercase{\textit{et al.}}: \papertitle}

\maketitle

\begin{abstract}
Windows malware detectors based on machine learning are vulnerable to adversarial examples, even if the attacker is only given black-box query access to the model.
The main drawback of these attacks is that: ($i$) they are query-inefficient, as they rely on iteratively applying random transformations to the input malware; and ($ii$) they may also require executing the adversarial malware in a sandbox at each iteration of the optimization process, to ensure that its intrusive functionality is preserved.
In this paper, we overcome these issues by presenting a novel family of black-box attacks that are both query-efficient and functionality-preserving, as they rely on the injection of benign content (which will never be executed) either at the end of the malicious file, or within some newly-created sections. 
Our attacks are formalized as a constrained minimization problem which also enables optimizing the trade-off between the probability of evading detection and the size of the injected payload.  
We empirically investigate this trade-off on two popular static Windows malware detectors, and show that our black-box attacks can bypass them with only few queries and small payloads, even when they only return the predicted labels.
We also evaluate whether our attacks transfer to other commercial antivirus solutions, and surprisingly find that they can evade, on average, more than 12 commercial antivirus engines.
We conclude by discussing the limitations of our approach, and its possible future extensions to target malware classifiers based on dynamic analysis.
\end{abstract}


%
\IEEEpeerreviewmaketitle

\section{Introduction}

\lettrine{M}{achine learning} is becoming ubiquitous in the field of computer security.
Both academia and industry are investing time, money and human resources to apply these statistical techniques to solve the daunting task of malware detection.
In particular, Windows malware is still a threat in the wild, as thousands of malicious programs are uploaded to VirusTotal every day.\footnote{\url{https://www.virustotal.com/it/statistics/}}
Modern approaches use machine learning to detect such threats at scale, leveraging many different learning algorithms and feature sets~\cite{saxe2015deep,kolosnjaji2016deep,hardy2016dl4md,david2015deepsign,incer2018adversarially, anderson2018ember, raff2017malware}.

While these techniques have shown promising malware-detection capabilities, they have not been originally designed to deal with non-stationary, adversarial problems in which attackers can manipulate the input data to evade detection.
This has been widely shown in the last decade in the area of \emph{adversarial machine learning}~\cite{huang2011adversarial,biggio18}. 
This research field studies the security aspects of machine-learning algorithms under attacks staged either at training or at test time.
In particular, in the context of learning-based Windows malware detectors, it has been shown that it is possible to carefully optimize \emph{adversarial malware} samples against the target system to bypass it~\cite{demetrio2019explaining, kolosnjaji2018adversarial,kreuk2018adversarial, castro2019armed, labaca-castro2019aimed, anderson2017evading, rosenberg2018generic, hu2017generating}.
Many of these attacks have been demonstrated in the black-box setting in which the attacker has only query access to the target model~\cite{labaca-castro2019aimed, anderson2017evading,rosenberg2018generic,hu2017generating}.
This really questions the security of such systems when deployed as \emph{cloud services}, as they can be queried by external attackers who can in turn optimize their manipulations based on the feedback provided by the target system, until evasion is achieved. 

These black-box attacks are however still not very efficient in terms of ($i$) the number of required queries, ($ii$) the complexity of their optimization process, and ($iii$) the amount of manipulations performed on the input sample, as detailed below.
First, query efficiency is hindered by the fact that these attacks optimize the adversarial malware by iteratively applying transformations which are not specifically targeted to evade detection, like injection of \emph{random} bytes after the end of the file. 
Second, the optimization process may be quite computationally demanding as some attacks require executing the adversarial malware sample in a sandbox at each iteration to ensure that its intrusive functionality is preserved. 
This verification step is required by attacks that either manipulate data in feature space (rather than considering realizable input modifications~\cite{tong2019improving}), or consider input transformations that may break the functionality of the malware sample~\cite{xu2016automatically,labaca-castro2019aimed}.
While executing the malware sample once inside a sandbox may not significantly slow down the whole process, the problem becomes relevant when this step has to be repeated after each iteration of the optimization process, as it requires restoring the state of the virtual environment at the stage before infection.
In addition, many malware samples can detect if they are run in a virtual environment and delay their execution to stay undetected~\cite{afianian2019malware}. 
This makes the problem of verifying that malware functionality is preserved even more complicated.
%
Third, all these attacks achieve evasion by significantly manipulating the content of the input malware, without considering additional constraints, \eg, on the resulting file size or number of injected sections. This may result in attack samples that are easily detected as anomalous by only looking at some trivial characteristics, like the file size or the number of sections. 

%

In this paper, we aim to overcome the aforementioned limitations by proposing a novel family of black-box attacks (Sect.~\ref{sec:black-box}) that can efficiently optimize adversarial malware samples.
First, our attacks are \emph{query-efficient}, as they rely upon injecting content specifically targeted to facilitate evasion, \ie, extracted from \emph{benign} samples (instead of being randomly generated).
Second, they are \emph{functionality-preserving} by design, as they leverage a set of manipulations that only inject content into the malicious program by exploiting ambiguities of the file format used to store programs on disk, without altering its execution traces. While in this work we only focus on injecting content either  at  the  end  of  the  file (\emph{padding})  or  within  some  newly-created  sections  (\emph{section injection}), our approach is general enough to encompass a wider range of functionality-preserving manipulations (as discussed in Sect.~\ref{subsec:realizable-attacks}).
Finally, our attacks are \emph{stealthier}. In particular, they are formalized as a constrained minimization problem which does not only optimize the probability of evading detection, but also penalizes the size of the injected adversarial payload via a specific regularization term.

We focus on two popular learning-based Windows malware detectors, built on features extracted from static code analysis (Sect.~\ref{sec:windows}). 
Our empirical evaluation (Sect.~\ref{sec:experiments}) investigates the trade-off between detection and size empirically, and shows that our black-box attacks are able to efficiently bypass the considered detectors after only few iterations and changes.
Moreover, we show that our attacks succeed not only when the target models output a continuous probability (or confidence) score, but also when they only provide the predicted labels.
We then evaluate whether our attacks transfer to other commercial antivirus solutions, and surprisingly find that they  can  evade,  on  average,  more than  12  commercial antivirus  engines.
We discuss how related work differs from ours in Sect.~\ref{sec:related}, and acknowledge the limitations of our work in Sect.~\ref{sec:limitations}.
We conclude by discussing possible future extensions of this work (Sect.~\ref{sec:conclusion}), including how to extend it to target malware classifiers based on dynamic analysis.
\section{Programs and Malware Detection}
\label{sec:windows}

In this section we first discuss the Windows \emph{Portable Executable} (PE) format,\footnote{\url{https://docs.microsoft.com/en-us/windows/win32/debug/pe-format}}
which describes how programs are stored on disk, and explains to the operating system (OS) how to load them in memory before execution. We then introduce the two popular learning-based Windows malware detectors used in the remainder of this work.

\subsection{The Windows Portable Executable (PE) File Format}
\begin{figure}[t!]
    \centering
    \includegraphics[width=0.45\textwidth]{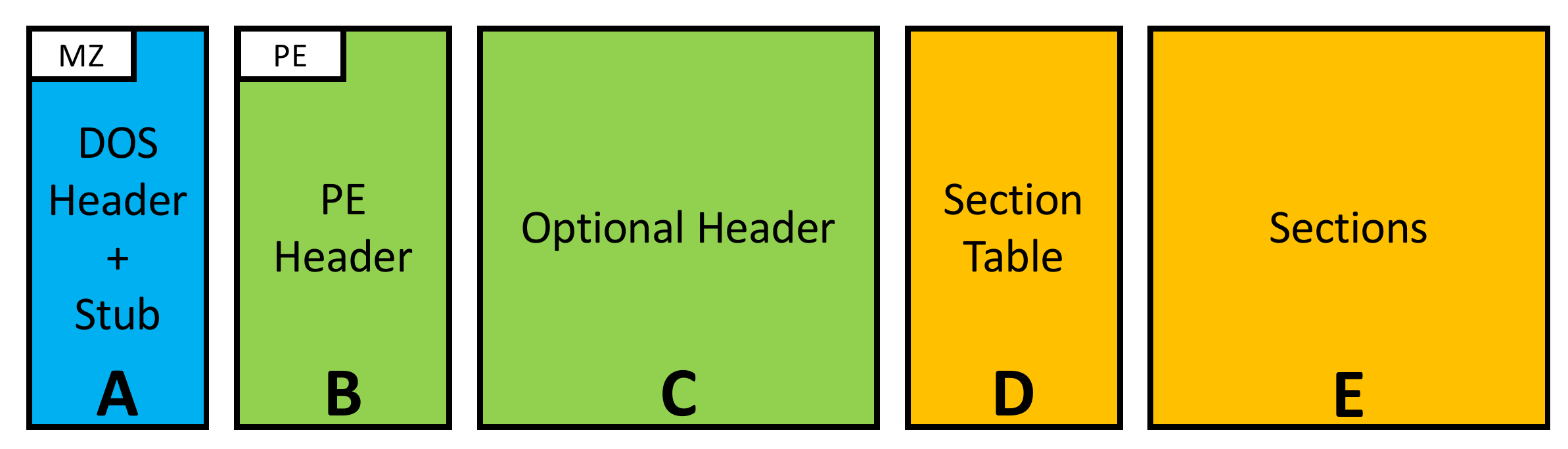}
    \caption[]{The Windows PE file format.
    Each colored section describes a particular characteristic of the program.
    }
    \label{fig:pe_format}
\end{figure}

The Windows PE format consists of several components, as shown in Fig.~\ref{fig:pe_format} and described below.

\myparagraph{DOS Header (A).} It contains metadata for loading the executable inside a DOS environment, and the \emph{DOS stub}, which prints ``\emph{This program cannot be run in DOS mode}'' if executed inside a DOS environment. 
These two components have been kept to maintain compatibility with older Microsoft's operating system. 
From the perspective of a modern application, the only relevant portions present inside the DOS Header are: (i) the magic number \verb|MZ|, a two-byte long signature for the file, and (ii) the four-byte long integer at offset \verb|0x3c|, that works as a pointer to the real header. 
If one of these two values is scrambled for some reason, the program is considered corrupted, and it will not be executed by the OS.

\myparagraph{PE Header (B).} It contains the magic number \verb|PE| along with other file characteristics, such as the target architecture, the header size, and the file attributes.

\myparagraph{Optional Header (C).} It contains the information needed by the OS to initialize the loading program.
It also contains offsets that point to useful structures, like the Import Address Table (IAT), needed by the OS for resolving dependencies, and the Export Table offset, which indicates where to find functions that can be referenced by other programs.

\myparagraph{Section Table (D).} It is a list of entries that indicates the characteristics of each core component of the program, and where the OS loader should find them inside the file.

\myparagraph{Sections (E).} These contiguous chunks of bytes host the real content of the executable. 
To list a few: \emph{.text} contains the code, \emph{.data} contains global variables and \emph{.rdata} contains read-only constants, and counting.

The structure of an executable program can be useful for statically inferring information about its behavior.
Indeed, most antivirus vendors apply static analysis to detect threats in the wild, without executing suspicious programs inside a controlled environment.
This approach saves time and resources since the antivirus programs do not execute the suspicious software inside the host OS.
Static analysis serves as the first line of defense, and its performance is crucial for opposing the countless threats in the wild.

\subsection{Learning-based Windows Malware Detection}

We focus on two popular, state-of-the-art machine learning-based detectors that have been coded, trained, and publicly released on GitHub by EndGame.\footnote{\url{https://www.endgame.com/}} Both models are trained on the EMBER dataset built by the same company~\cite{anderson2018ember}.

\myparagraph{MalConv.} The first detector is an end-to-end convolutional neural network (CNN) proposed by Raff et al.~\cite{raff2017malware}.
It takes as input the first 2 MB of an executable and returns the probability of being malware. 
If the input executable length exceeds this threshold, the file is truncated to the specified size, otherwise, the file is padded with the value \verb|0|.
Since the padding value should be unique, all values are shifted by one to maintain this distinction.
Each byte is embedded into a representation space with eight dimensions, learned directly from the available data with the goal of defining a meaningful distance metric between bytes. The convolutional layers are then used to correlate spatially-distant bytes inside the input binary, \eg, jumps and function calls.

\myparagraph{GBDT.} The second detector is implemented using Gradient Boosting Decision Trees (GBDT)~\cite{anderson2018ember,ke2017lightgbm}.
Differently from MalConv, this detector uses a fixed representation consisting of 2,381 features, extracted from: 
($i$) \emph{general file information}, including the virtual size of the file, the number of imported and exported functions, the presence of debug sections, etc.;
($ii$) \emph{header information}, accounting for the characteristics of the executable, the target architecture, the version, etc.;
($iii$) \emph{byte histogram}, which counts the occurrences of each byte, divided by the total number of bytes;
($iv$) \emph{byte-entropy histogram}, inspired from~\cite{saxe2015deep}, which accounts for the entropy of the byte distribution of the file, applying a sliding window over the binary; 
($v$) \emph{information taken from strings}, which counts the number of occurrences of each string (considered as a sequence of at least five consecutive printable characters), and how many special markers they contain, such as \verb|C:\|, \verb|HKEY|, \verb|http| and \verb|https|;
($vi$) \emph{section information}, which includes name, length, entropy, and virtual size of each section;
($vii$) \emph{imported and exported functions}
which tracks all the functions imported from libraries, and all the ones that are exposed to the other programs.
Many of these feature sets are compressed inside a histogram by applying the \emph{hashing trick}~\cite{moody1989fast}, to reduce the dimension of the problem to a smaller and manageable space.

\section{Black-Box Optimization of Adversarial Windows Malware}
\label{sec:black-box}

In this section, we present our novel black-box attack framework, named \algoname~(Genetic Adversarial Machine learning Malware Attack).
\algoname{} can efficiently optimize adversarial malware samples while only requiring black-box access to the model, \ie, by only querying the target model and observing its output, without accessing its internal structure and parameters. 
Our attack relies upon a set of \emph{functionality-preserving} manipulations that inject content into the malicious program by exploiting ambiguities of the PE format used to store programs on disk, without altering its execution traces. This allow us to get rid of the computationally-demanding validation steps required to ensure that the manipulated malware preserves its intended functionality. 
In particular, we consider here content manipulations specifically targeted to facilitate evasion, \ie, extracted from \emph{benign} samples rather than being generated at random. While this makes our attack much more \emph{query-efficient}, it is worth remarking that our framework is general enough to encompass many other different content manipulation techniques, as detailed in Sect.~\ref{subsec:realizable-attacks}.
Finally, to make our attack \emph{stealthier}, we formalize it as a constrained optimization problem which does not only minimize the probability of evading detection but also the size of the injected content via a specific penalty term. 

\begin{figure*}
    \centering
    \includegraphics[width=0.95\textwidth]{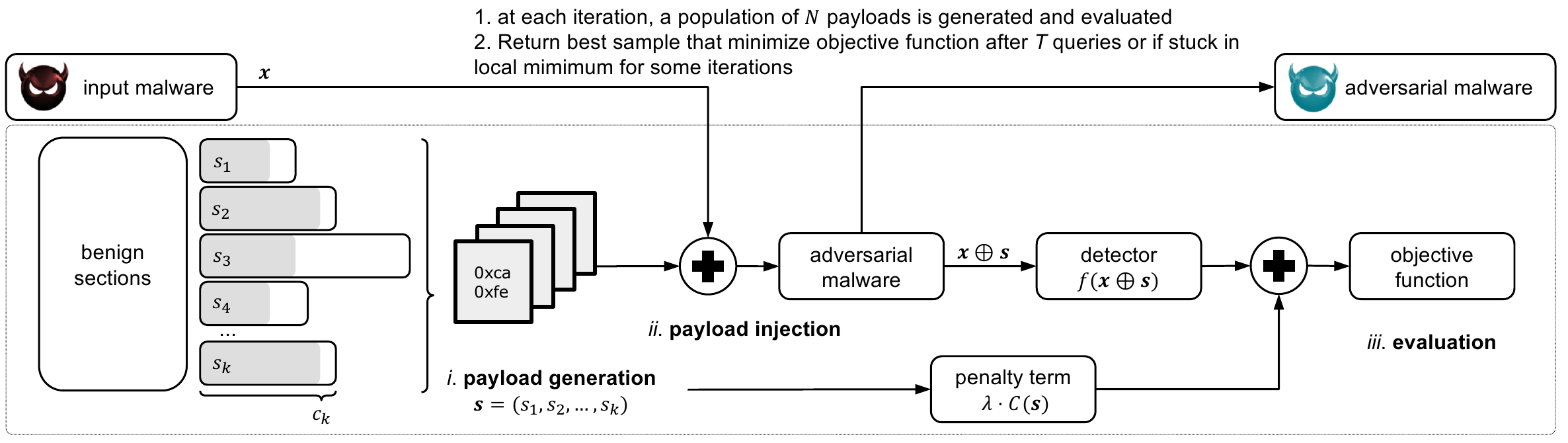}
    \caption[]{
    Conceptual schema of \algoname.
    The optimizer (i) generates different payload extracted from benign programs, (ii) injects such content inside the original malware, and (iii) computes the objective function, by combining the response of the detector and the size constraint controlled by $\lambda$.
    The process ends after \emph{T} queries, or if it converges earlier, \ie, if the objective function does not significantly decrease after a given number of iterations.}
    \label{fig:algo}
\end{figure*}

\myparagraph{Notation.} In the following, we denote with $\vct x \in \programs \subset \{0, \dots , 255\}^*$ the (malicious) input program, described as a string of bytes of arbitrary length. 
We then define a set of $k$ distinct functionality-preserving manipulations that can be applied to the input program $\vct x$ as a vector $\vct s \in \faspace \subset [0,1]^k$.
Each element of $\vct s$ corresponds to a different manipulation that can be applied to the input program. The manipulations are parameterized in $[0,1]$, to denote the extent to which they are applied. For example, if we assume that the $i^{\rm th}$ element $s_i$ is associated to the injection of a given section in the input program, $s_i = 0.4$ may represent the fact that only 40\% of the bytes present in that section will be injected. We can also consider injection of specific API functions, in which case $s_i$ will be a binary variable denoting whether the given API is injected ($s_i=1$) or not ($s_i=0$).
The function $\oplus : \programs \times \faspace \rightarrow \programs$ applies the manipulations described by $\vct s$ to the input program $\vct x$, preserving functionality, and returns the manipulated program.
We use $f : \programs \rightarrow \mathbb R$ to denote the output of the classification model on the input program. Without loss of generality, we consider here $f$ to be the output of the model on the malicious class, \ie, the higher the value of $f(\vct x)$ is, the more $\vct x$ is considered malicious.
The value of $f(\vct x)$ is eventually compared against a decision threshold $\theta$ to decide whether the input program is malicious, \ie, $f(\vct x) \geq \theta$, or not.

\myparagraph{Attack formulation.} We can now formalize our attack as the following constrained minimization problem:
\begin{mini}|l|
    {\vct s \in \faspace}{F(\vct s) = f(\vct x \oplus \vct s) + \lambda \cdot \size(\vct s)}{}{} \, ,
    \addConstraint{\query}{\leq T}{} \, .
    \label{eq:minimizer}
\end{mini}
The objective function $F(\vct s)$ consists of two conflicting terms: ($i$) $f(\vct x \oplus \vct s)$, \ie, the classification output on the manipulated program, and ($ii$) $\size(\vct s)$, \ie, a penalty function that evaluates the number of injected bytes into the input malware. 
The hyperparameter $\lambda > 0$ tunes the trade-off between these two terms, \ie, it promotes solutions with smaller number of injected bytes $\size(\vct s)$ at the expense of reducing the probability that the sample is misclassified as benign (larger $f(\vct x \oplus \vct s)$ values). Varying the hyperparameter $\lambda$ allows us to evaluate how the attack effectiveness increases as a function of the size of the injected adversarial payload.

The objective $F$ is minimized w.r.t. the choice of the applied manipulations $\vct s$. In this work, we restrict the available manipulations $\vct s=(s_1, \ldots, s_k)$ to the injection of content extracted from a predefined set of $k$ benign sections, without optimizing the content-injection location. This means that $s_i$ will represent the fraction of bytes extracted from the $i^{\rm th}$ benign section, and these bytes will be injected before those extracted from section $s_j$, for $j > i$. 
In this context, we define the penalty term $\size(\vct s)$ as $\size(\vct s) = \vct c ^T \vct s$, where $\vct c \in \mathbb R^k$ is a vector whose $i^{\rm th}$ element $c_i$ is equal to the overall size of the $i^{\rm th}$ benign section. Accordingly, $\size(\vct s)$ measures the size of the injected payload.
As the elements of $\vct c$ and $\vct s$ are non-negative, this term penalizes a weighted version of the $\ell_1$ norm of $\vct s$, thus promoting sparse solutions. This means that many elements of the optimal solution vector $\vct s^\star$ will be zero, \ie, only content from few benign sections will be injected.

As optimizing the objective in a black-box manner requires querying the target model $f$ repeatedly, we use the constraint $q \leq T$ to upper bound the maximum number of queries $q$ that can be performed by $T$. The query budget $T$ is another hyperparameter in our approach, and increasing it allows our attack to better optimize the trade-off between misclassification confidence and payload size, at the expense of an increased computational complexity.

\myparagraph{Solution algorithm.} We solve the given minimization problem using a black-box genetic optimizer, as detailed in Algorithm~\ref{fig:genetic_algorithm} and Fig.~\ref{fig:algo}. 
The algorithm is initialized by randomly generating a matrix $\mat S^\prime = (\vct s_{1}, ..., \vct s_{N}) \in \faspace^N \subset [0,1]^{N \times k}$, which represents the initial population of $N$ candidate manipulation vectors (line \ref{line:initial_population}). 
Then, the genetic algorithm iterates over three steps that mimic the process of biological evolution: \emph{selection}, \emph{cross-over}, and \emph{mutation}.
The \emph{selection} step (line \ref{line:selection}) uses the objective function to evaluate the candidates in $\mat S^\prime$, and selects the best $N$ candidates between the current population $\mat S^\prime$ and the population generated at the previous iteration $\mat S$. These are the candidate manipulation vectors associated with the lowest values of $F$.
The \emph{crossover} function (line \ref{line:crossover}) takes the selected candidates as input and returns a novel set of $N$ candidates by mixing the values of pairs of randomly-chosen vector candidates. In particular, given a pair of candidate vectors from the previous population, a new candidate is generated by cloning the values $s_1, \ldots, s_j$ from the first parent and the remaining values $s_{j+1}, \ldots, s_k$ from the second parent, being $j \in \{1,\ldots, k\}$ an index selected at random.
The \emph{mutation} function (line \ref{line:mutation}) changes the elements of each input vector at random, with low probability. The combination of both \emph{cross-over} and \emph{mutation} ensures that the new population is sufficiently different from the previous one, allowing the algorithm to properly explore the space of feasible solutions.

In each iteration, the algorithm performs $N$ new queries to the target model, to evaluate the objective $F$ on the new candidates in $\mat S^\prime$, and then retains the best candidate population $\mat S$. When either the maximum number of queries $T$ or convergence is reached (\eg, if no further improvement in the value of $F$ is observed across a given number of iterations), the algorithm returns the best manipulation vector $\vct s^\star$ from the current population $\mat S$.
The corresponding optimal adversarial malware $\vct x^\star$ can be finally obtained by applying the optimal manipulation vector $\vct s^\star$ to the input sample $\vct x$ through the manipulation operator $\oplus$ as $\vct x^\star = \vct x \oplus \vct s^\star$.

\begin{algorithm}[t]
    \SetKwInOut{Input}{Input}
    \SetKwInOut{Output}{Output}
    \Input{$\vct x$, the initial malware sample; $\lambda$, the regularization parameter; $N$, the population size; $T$, the query budget.}
    \Output{$\vct s^\star$, the manipulations which minimize $F$.}
    $q \leftarrow 0$, $\mat S \leftarrow \emptyset$ \\
    $\mat S^\prime \leftarrow (\vct s_1, ..., \vct s_N) \in \faspace^N$ \label{line:initial_population}\\
    \While{ $ q < T$ \textbf{and} not converged }{
    \label{algo:loop}
        $\mat S \leftarrow \text{selection}(\mat S \cup \mat S^\prime, F, \vct x, \lambda)$ \label{line:selection}\\
        $\mat S^\prime \leftarrow$ crossover($\mat S$)\label{line:crossover}\\
        $\mat S ^\prime\leftarrow$  mutate($\mat S^\prime$) 
        \label{line:mutation}\\
        $q \leftarrow q + N$\\
    }
    \textbf{return} $\vct s^\star$, best candidate from $\mat S$ with minimum $F$.
    \caption{Genetic optimization of adversarial malware with \algoname.}
    \label{fig:genetic_algorithm}
\end{algorithm}

\myparagraph{Hard-label attacks.} In some cases, the target model may only provide the classification label assigned to the input sample, instead of a continuous confidence value $f(\vct x)$. In this hard-label scenario, we adapt \algoname by setting $f(\vct x)=0$ if the input sample is classified as benign, and $f(\vct x)=\infty$ otherwise, to discard perturbed malware samples that do not evade detection. 
This does not substantially hinder the success of our algorithm, as evasive malware variants can be typically found by injecting a sufficiently-large amount of benign content into the initial malware. Once an evasive malware variant is found, the genetic algorithm starts reducing the injected payload size $\size(\vct s)$ iteratively, while trying to preserve misclassification.

\subsection{Functionality-preserving manipulations}
\label{subsec:realizable-attacks}

We discuss here the set of functionality-preserving manipulations that can be used in our attack framework.
In the context of Windows PE file format, there are only a few transformations that can be applied without compromising the execution of the input program. We categorize them either as structural or behavioral, as detailed below.

\myparagraph{Structural.} This family of manipulations affects only the structure of the input program, by exploiting ambiguities inside the file format, without altering its behaviour.

\noindent \textit{($s.1$) Perturb Header Fields}~\cite{anderson2017evading, labaca-castro2019aimed, castro2019armed}. This technique includes altering section names, breaking the checksum, and altering debug information. These are fine-grained manipulations that can be applied to the PE components \emph{B}, \emph{C} and \emph{D} in Fig.~\ref{fig:pe_format}.

\noindent \textit{($s.2$) Filling Slack Space}~\cite{kreuk2018adversarial, suciu2019exploring, anderson2017evading, castro2019armed, labaca-castro2019aimed}. This technique manipulates the slack space inserted by the compiler to maintain the alignments inside the file. The corresponding slack bytes (inside \emph{E} in Fig.~\ref{fig:pe_format}) are usually set to zero, and they are never referenced by the code of the executable.

\noindent \textit{($s.3$) Padding}~\cite{kolosnjaji2018adversarial, suciu2019exploring, kreuk2018adversarial}. This technique injects additional bytes at the end of the file (after \emph{E} in Fig.~\ref{fig:pe_format}).

\noindent \textit{($s.4$) Manipulating DOS Header and Stub}~\cite{demetrio2019explaining, demetrio2020adversarial}. This technique modifies some  bytes in the DOS Header (\emph{A} in Fig.~\ref{fig:pe_format}) which are not used by modern programs (see Sect.~\ref{sec:windows}).

\noindent \textit{($s.5$) Extend the DOS Header}~\cite{demetrio2020adversarial}. This technique extends the DOS header by injecting content before the actual header of the program (between \emph{A} and \emph{B} in Fig.~\ref{fig:pe_format}).

\noindent \textit{($s.6$) Content shifting}~\cite{demetrio2020adversarial}. This technique creates additional space before the beginning of a section, by shifting the content forward, and injects adversarial content in between (\ie, between \emph{D} and \emph{E} in Fig.~\ref{fig:pe_format}).

\noindent \textit{($s.7$) Import Function Injection}~\cite{anderson2017evading, labaca-castro2019aimed, castro2019armed}. This technique injects import functions by adding an appropriate entry to the Import Address Table, specifying which function from which library must be included during the loading process (this affects \emph{C} and \emph{E} in Fig.~\ref{fig:pe_format}). 

\noindent \textit{($s.8$) Section Injection}~\cite{anderson2017evading, labaca-castro2019aimed, castro2019armed}. This technique injects new sections into the input file by creating an additional entry inside the section table (thus affecting \emph{D} and \emph{E} in Fig.~\ref{fig:pe_format}).
Each section entry is 40 bytes long, so all the content has to be shifted by that amount, without compromising file and section alignments as specified by the header.

\myparagraph{Behavioral.} This family of perturbations can change the program behavior and execution traces, but still preserving the intended functionality of the malware program. For example, these transformations encompass the binary rewriting techniques in~\cite{wenzl2019hack}, as  discussed below.

\noindent \textit{($b.1$) Packing}~\cite{anderson2017evading, labaca-castro2019aimed, castro2019armed}. This technique amounts to encrypting or encoding the content of the binary inside another binary and decoding it at run-time. The effect of a packer is invasive since the whole structure of the input sample is modified.

\noindent \textit{($b.2$) Direct}~\cite{wenzl2019hack}. This approach rewrites specific portions of the code, like replacing assembly instructions with equivalent ones (\eg, additions and subtractions with opposed sign).

\noindent \textit{($b.3$) Minimal Invasive}~\cite{anderson2017evading,wenzl2019hack}. This technique sets the entry-point to a new executable section that jumps back to the original code.

\noindent \textit{($b.4$) Full Translation}~\cite{wenzl2019hack}. This approach lifts all the code to a higher representation, \eg, LLVM,~\footnote{\url{https://llvm.org/}} since it simplifies the application of perturbations, and it then translates the code back to the assembly language.

\noindent \textit{($b.5$) Dropper}~\cite{ceschin2019shallow}. This approach stores the code as a resource of another binary, which is then loaded at runtime.

\myparagraph{Padding and section-injection attacks.} In this work, we implement \algoname using two different structural manipulation techniques, \ie, \emph{padding} and \emph{section injection}, and refer to them respectively as \emph{padding} and \emph{section-injection} attacks.
The rationale behind this choice is related to the complementary nature of these transformations. Padding is the easiest manipulation that can be applied and, similarly to $s.1 - s.6$, it injects content into the unused space of the executable, without altering its structure. Section injection, instead, does not only allow injecting custom content like the other techniques, but it also manipulates the structure of the executable by adding a section entry inside the section table. 

\section{Experiments}
\label{sec:experiments}
In this section, we empirically evaluate the effectiveness of our attacks against both the GBDT and MalConv malware detectors. We ran our experiments on a workstation equipped with an Intel\textsuperscript{\textregistered}
 Xeon\textsuperscript{\textregistered}
 CPU E5-2670, with 48 CPU and 128 GB of RAM.
The pre-trained version of MalConv presents a slightly different architecture w.r.t. the original formulation: 1 MB of input size and padding value of \verb|256| to avoid the shifting pre-processing part.
The network is implemented using PyTorch~\cite{paszke2019pytorch}.
We developed the genetic optimizer of \algoname using \emph{DEAP}\cite{DEAP_JMLR2012}.
We tested the attack using a population size $N$ of $10$ elements, varying the query budget $T$ from 10 to 510.
If the optimizer stagnates in a local minimum for more than 5 iterations, we halt the process.
We used values for the regularization parameter $\lambda \in \{10^{-i}\}_{i=3}^{9}$.
Since the attack feature space $\faspace$ is parametric over the number of sections the attacker may add, we randomly extract 75 \verb|.rdata| sections from our goodware dataset that will be used for adding content to the input malware, for a maximum of 2.5 MB, as discussed in Section \ref{subsec:realizable-attacks}.
We willingly set this number high, as the optimizer will find small payloads thanks to the sparsity imposed by the penalization term that behaves as a $\ell_1$ norm.
We implemented and open-sourced the library we used for computing these attacks, named \texttt{secml-malware}.\footnote{\url{https://github.com/zangobot/secml_malware}}

\myparagraph{Performance in the absence of attack.}
\label{subsec:roc}
\begin{figure}[t!]
    \centering
    \includegraphics[width=0.8\linewidth]{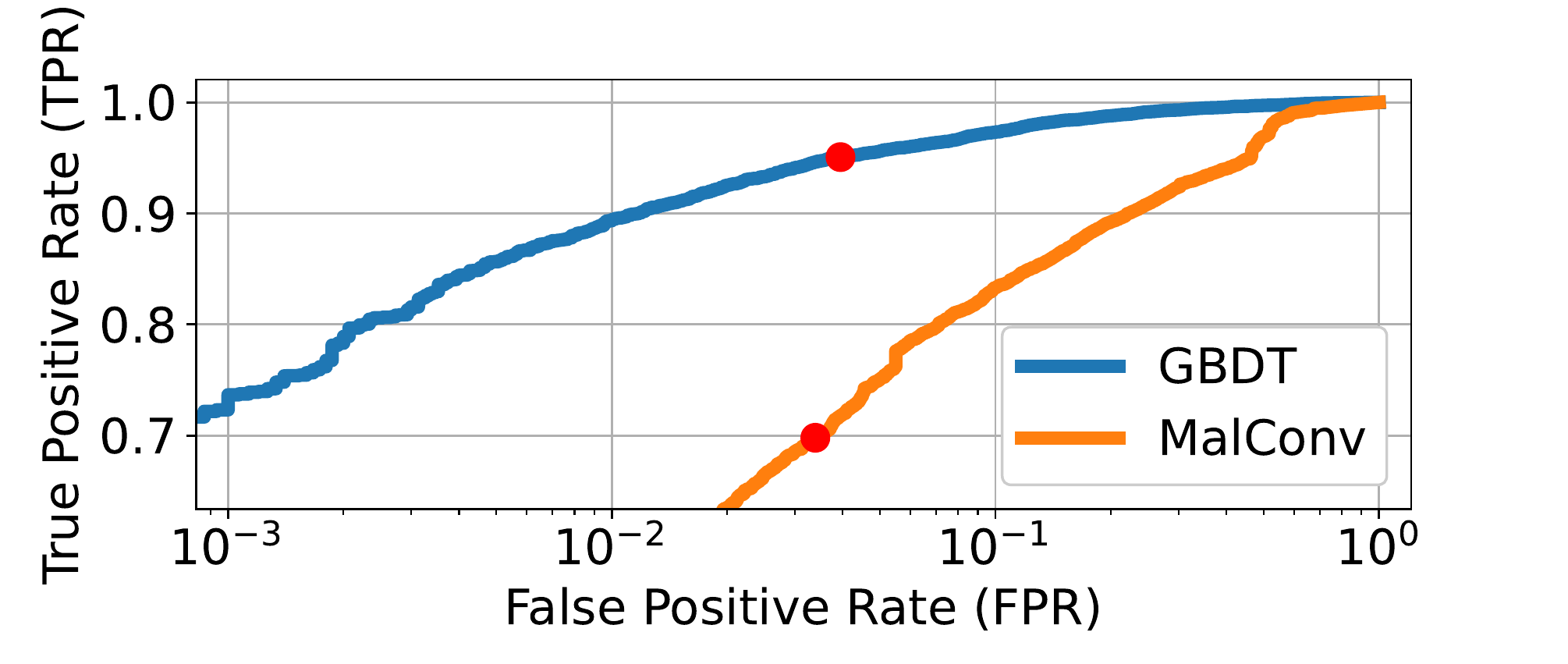}
    \caption{Receiver Operating Characteristic (ROC) curve of both classifiers.}
    \label{fig:roc}
\end{figure}
To evaluate the performance of both classifiers in the absence of attacks, we collected a set of $15,000$ benign and $15,000$ malware samples.
\begin{figure*}[t!]
	\centering
	\includegraphics[width=0.95\textwidth]{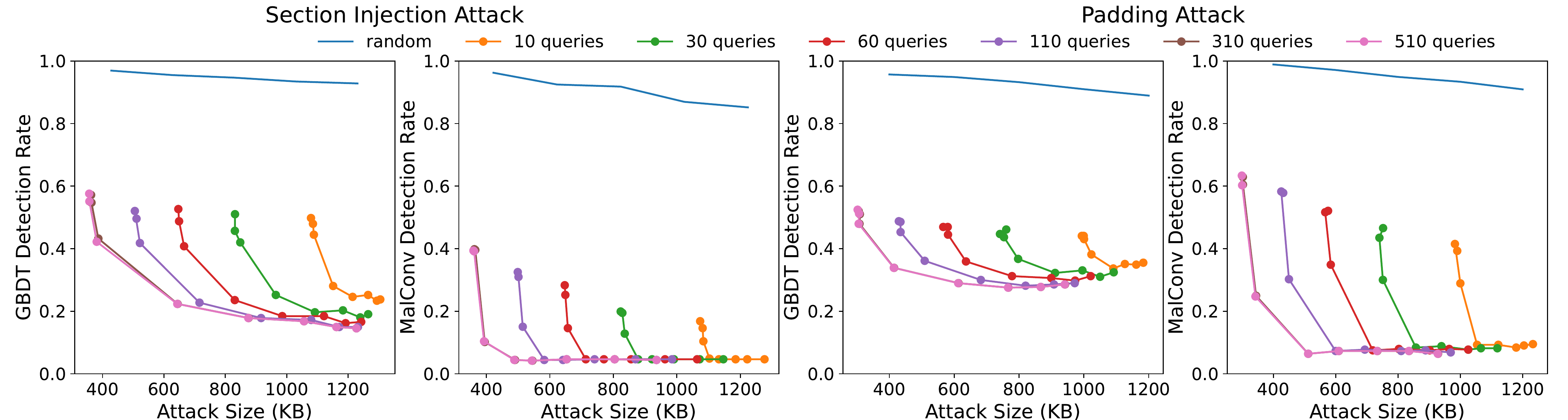}
\caption{Padding and section injection attack performances for $\lambda \in \{10^{-i}\}_{i=3}^{9}$, using \nmalwaretest malware samples as input.
	The solid lines are computed as a regression over the point of a particular setting of the experiments.}
\label{fig:attack_reg_effect}
\end{figure*}
The malware samples were gathered from VirusTotal,\footnote{\url{https://www.virustotal.com}} while
the goodware samples were collected by downloading executable programs from GitHub.
The results are shown in Figure~\ref{fig:roc}.
The threshold chosen for GBDT is 0.8336, which corresponds to an False Positive Rate (FPR) of 0.039 and a True Positive Rate (TPR) of 0.95.
The threshold used for MalConv is 0.5, that lead to a FPR of 0.035 and a TPR of 0.69.
The red dots inside the plot shows such values directly on the curve.
These results are comparable to the description given by the authors of GBDT~\cite{anderson2018ember}, as both detectors achieve just a slightly lower score \wrt what is reported in the paper.
Still, they can be both used as a baseline for our analysis.

\myparagraph{Attack evaluation.} We randomly sample \nmalwaretest from our collection of 15K malware set to use during the adversarial attacks, and this set includes 5.3\% ransomware, 29\% downloaders, 18\% viruses, 7\% backdoors, 29\% grayware, 8\% worms, plus other families with lower percentage.
Figure~\ref{fig:attack_reg_effect} shows how both the detection rate and adversarial payloads size vary \wrt the number of queries and the value of the regularization parameter.
Each curve in the plot has been produced by computing the mean detection rate and mean size for each values of $\lambda$, repeated
for different numbers of queries sent.
As the value of $\lambda$ decreases, the algorithm finds more evasive samples with bigger payloads, since the penalty term is negligible while computing the objective function.
On the other hand, by increasing the value of $\lambda$ the resulting attack feature vector become sparse, generating smaller but more detectable adversarial example.
In this case, the penalty term engulfs the score computed by the classifier, which becomes irrelevant during the optimization.
Another significant effect is posed by the number of total queries used by the genetic optimizer: the more are sent, the better the adversarial examples are in both detection rate and size.
Intuitively, by sending more queries, \algoname can explore more solutions that are stealthy and evasive at the same time, but such solutions could not be found at early stages of the optimization process.
To prove the efficacy of our methodology, we report the results of the application of random byte sequences of increasing length.
This experiment highlights a slight descending trend, but the optimized attack with benign content injection is way more effective than random perturbations.
The detection rate of GBDT is decreased more by the section-injection attack than by padding.
Since the first technique also introduces a section entry inside the section table, the adversarial payload perturbs more features than those modified by the padding attack.

\myparagraph{Hard-label attacks.}
We show aggregate results in Table \ref{tab:hard_vs_soft}, highlighting the comparisons between the performances of the soft-label and hard-label attacks.
Each entry presents the mean detection rate and the mean adversarial payload size for each detector, given a pair of number of queries / regularization parameter used for computing the specified attack.
We computed 4 different values of $\lambda$ in the set $\{10^{-(2i+1)}\}_{i=1}^4$.
Results suggest that, without the confidence score, once one evasive payload is found, then its size is optimized iteration after iteration of the genetic algorithm, regardless of the value of the regularization parameter $\lambda$.
This is caused by the settings we impose for our experiments: we used an infinite value to discard each detected adversarial example, hence all the remaining ones are used for optimizing only the size, acting as a constraint itself for the optimization.
The effectiveness of our methodology in this setting is caused by the nature of the content injected, that mimics the benign class, and this is confirmed by Figure \ref{fig:attack_reg_effect} where injecting random byte sequences has no effect against the targets.
On the contrary, the number of queries serves itself as a regularizer, since too few queries lead to larger adversarial payloads with low confidence, and numerous queries led to small payload whose score is higher.

\begin{table*}[]
\centering
{%
\begin{tabular}{lcccccc}
 &
  \multicolumn{2}{c}{\textbf{Soft-Label GAMMA (Q = 500)}} &
  \multicolumn{2}{c}{\textbf{Hard-Label GAMMA (Q = 30)}} &
  \multicolumn{2}{c}{\textbf{Hard-Label GAMMA (Q = 500)}} \\ \hline
\textbf{Padding} &
  \textit{GBDT} &
  \textit{MalConv} &
  \textit{GBDT} &
  \textit{MalConv} &
  \textit{GBDT} &
  \textit{MalConv} \\ \hline
$\lambda=10^{-9}$ & 28\% / 941 KB    & 6\% / 927 KB  &                &              &               &              \\ \cline{1-3}
$\lambda=10^{-6}$ & 33\% / 413 KB    & 6\% / 511KB   & 35\% / 945 KB  & 6\% / 835 KB & 30\% / 661 KB & 5\% / 473 KB \\ \cline{1-3}
$\lambda=10^{-3}$ & 52\% / 302 KB & 63\% / 298 KB &                &              &               &              \\ \hline \hline
\vspace{0.5pt}\textbf{Section Injection} &
  \textit{GBDT} &
  \textit{MalConv} &
  \textit{GBDT} &
  \textit{MalConv} &
  \textit{GBDT} &
  \textit{MalConv} \\ \hline
$\lambda=10^{-9}$ & 14\% / 1227 KB   & 4\% / 935 KB  &                &              &               &              \\ \cline{1-3}
$\lambda=10^{-6}$ & 22\% / 643 KB    & 10\% 487 KB   & 31\% / 1080 KB & 4\% / 880 KB & 21\% / 830 KB & 4\% / 490KB  \\ \cline{1-3}
$\lambda=10^{-3}$ & 57\% / 356 KB    & 39\% / 359 KB &                &              &               &              \\ \hline
\end{tabular}}
\caption{Comparison of soft-label and hard-label attacks, with different number of queries sent and values of $\lambda$.}
\label{tab:hard_vs_soft}
\end{table*}

\myparagraph{Temporal analysis.}
\begin{table}[h!]
\centering
\begin{tabular}{lcc}
                           & \textbf{GBDT} & \textbf{MalConv} \\ \hline
\textbf{Padding}           & 0.60 $\pm$ 0.24s        & 0.93 $\pm$ 0.33s         \\ \hline
\textbf{Section injection} & 0.60 $\pm$ 0.30s        & 0.86 $\pm$ 0.20s         \\ \hline
\end{tabular}
\caption{Mean elapsed time for each attack and target.}
\label{tab:time}
\end{table}
From a temporal point of view, the complexity of \algoname~ is dominated by the time spent querying the detector.
Table~\ref{tab:time} shows the mean elapsed time needed to compute one single query, for each attack and target.
Surprisingly, the sum of the time spent by the feature extraction phase and the prediction of GBDT is less than the time needed by the neural network to process all the bytes.

\myparagraph{Packing effect.}
\begin{figure}[t]
\centering
    \includegraphics[width=0.45\textwidth]{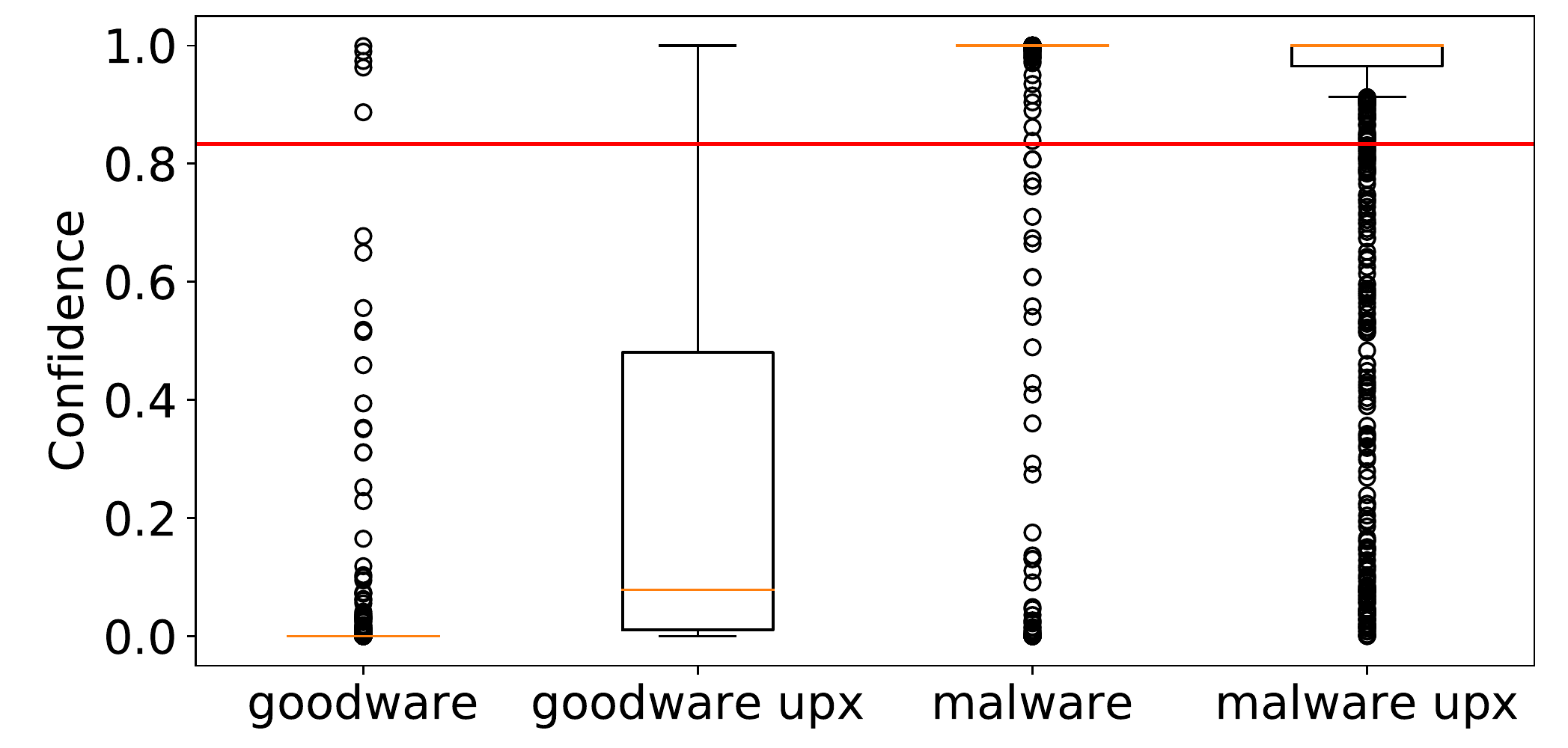}
    \includegraphics[width=0.45\textwidth]{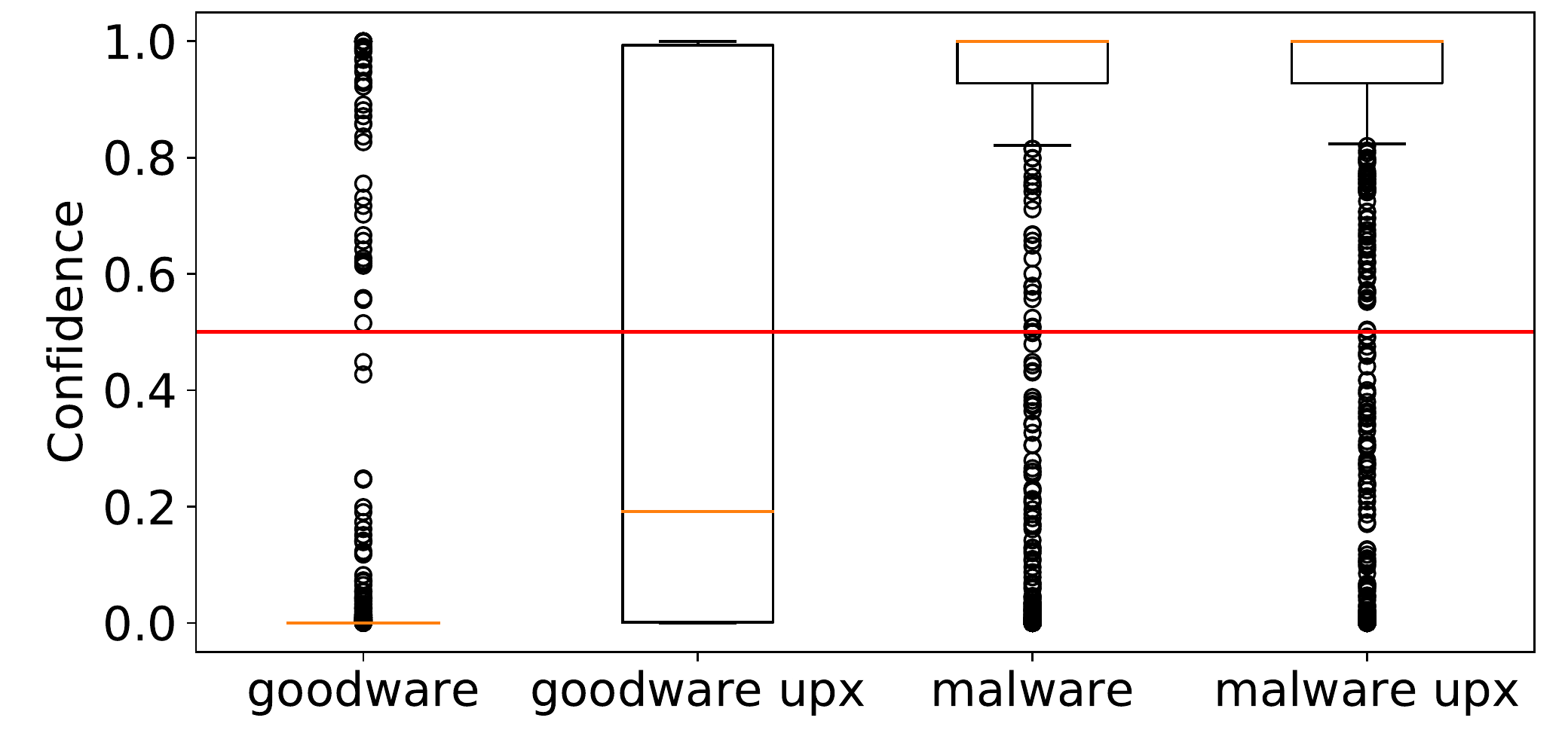}
    \caption{Effect of UPX packing on GBDT (top) and MalConv (bottom). 
    Each box-plot shows the distribution of the classifier confidence $f(\vct x)$ on the malicious class, and the dashed red line is the decision threshold $\theta$.} 
    \label{fig:packing}
\end{figure}
Since these classifiers leverage only static features, it is reasonable to ask ourselves whether encrypting the program content is already sufficient to evade detection, without applying all the techniques we have introduced in Section~\ref{sec:black-box}.
\emph{Packing} is a technique used to reduce the size of an executable, by applying a compression, encryption or encoding algorithm.
As the effect of a packer completely changes the program representation on disk, it has been extensively used by malware vendors to hide their product to the analysts, increasing the difficulty of reverse-engineering analyses.
In this context, we apply one famous technique, called UPX\footnote{\url{https://upx.github.io}} to 1000 malware and 1000 goodware programs, and test the evasion rate for both MalConv and GBDT.
The effectiveness of the UPX packer is shown in Figure~\ref{fig:packing}.
Both detectors attribute a malicious score when the sample is packed, and this is intuitive by looking at the box-plot of the packed goodware programs.
Both detectors increase their score towards the malware class, while there is only a little change in terms of mean and variance for packed malware.

From these results, we believe that application of packing techniques is seen as a malicious trait by the detector.
This might be caused by the abundance of packed malware inside the training set~\cite{anderson2018ember}, opposed to the scarcity of packed goodware.
As a result, models trained on such data might possess a bias that makes them wrongly assume that a sample is malicious only because it is packed.
Also, given enough samples packed with a technique, the learning algorithm should be able to capture the signature left by the packer itself inside the packed program.
For instance, the UPX packer creates two executable sections called \emph{UPX0} and \emph{UPX1}, that contain the extraction code and the original compressed program.
We believe that evasion through packing techniques should more likely to be achieved by unseen packers, \ie custom solutions developed by malware vendors themselves.

\begin{table}[]
\centering
{
\centering
\begin{tabular}{lccc}
    & \textbf{Malware}     & \textbf{Random} & \textbf{Sect. Injection} \\ \hline
\multicolumn{1}{c}{\textbf{Detections (VT)}} & 46.56 $\pm$ 12.40 & 40.80 $\pm$ 12.40     & 34.50 $\pm$ 12.63 \\ \hline
\end{tabular}}
\caption{Number of antivirus programs from VirusTotal (VT) that detect ($i$) the initial malware and its modified versions with ($ii$) random and ($iii$) section-injection attacks, averaged over \vttest{} malware samples (standard deviation is also reported). While random attacks evade 5.76 detectors, on average, section-injection attacks evade up to 12.01 detectors.}
\label{tab:vt_table}
\end{table}

\begin{table}[]
\centering
{
\centering
\begin{tabular}{lccc}
    & \textbf{Malware}     & \textbf{Random} & \textbf{Sect. Injection} \\ \hline
\multicolumn{1}{c}{\textbf{AV1}} & 93.5\%  & 85.5\% & 30.5\% \\ \hline 
\multicolumn{1}{c}{\textbf{AV2}} & 85.0\% & 78.0\% & 68.0\% \\ \hline 
\multicolumn{1}{c}{\textbf{AV3}} & 85.0\% & 46.0\% & 43.5\% \\ \hline 
\multicolumn{1}{c}{\textbf{AV4}} & 84.0\% & 83.5\% & 63.0\% \\ \hline  
\multicolumn{1}{c}{\textbf{AV5}} & 83.5\% & 79.0\% & 73.0\% \\ \hline  
\multicolumn{1}{c}{\textbf{AV6}} & 83.5\% & 82.5\% & 69.5\% \\ \hline 
\multicolumn{1}{c}{\textbf{AV7}} & 83.5\% & 54.5\% & 52.5\% \\ \hline  
\multicolumn{1}{c}{\textbf{AV8}} & 76.5\% & 71.5\% & 60.5\%  \\ \hline 
\multicolumn{1}{c}{\textbf{AV9}} & 67.0\% & 54.5\% & 16.5\% \\ \hline  
\end{tabular}}
\caption{Detection rate of 9 antivirus programs from VirusTotal computed on ($i$) the initial set of \vttest{} malware samples, and on the same samples manipulated with ($ii$) random attacks and ($iii$) section-injection attacks.}
\label{tab:vt_table2}
\end{table}

\myparagraph{Evaluation on antivirus programs (VirusTotal).} We evaluate here the impact of our attack on commercial detectors.
In this context, we are not interested in evading detection by these commercial programs, \eg, by packing the input samples, bur rather in assessing whether these methods can detect our attacks, given that our attacks only minimally modify the content of the input malware sample. In particular, the manipulations that we apply to our malware samples address only the syntactical structure of each program, and we aim to evaluate here if the application of such transformations can pose a threat to other antivirus programs.
We expect that most of the commercial solutions should not be affected by such attacks.
We rely on the response retrieved by VirusTotal,\footnote{\url{https://virustotal.com}} which is an online interface for many threat detectors.
The service offers an API that can be used for querying the system, by uploading samples from remote.
We test the performance of our attack by sending \vttest malware samples, before and after injecting the adversarial payload into the sample, optimized using the section-injection attack against the GBDT classifier.
We also compare our attack against a baseline random attack that simply adds a random payload of 50 KB to each sample.
Table~\ref{tab:vt_table} shows how many antivirus programs hosted on VirusTotal (70 in total) detect the submitted malware samples, on average. While the random attack only slightly decreases the number of detections per sample, the section-injection attack is able to bypass an average of more than 12 detectors per sample. 
To better evaluate the impact of our attack on individual antivirus programs, in Table~\ref{tab:vt_table2} we report the detection rates of 9 different antivirus products that appear on the 2019 Gartner Magic Quadrant for Endpoint Protection Platforms,\footnote{\url{https://www.microsoft.com/security/blog/2019/08/23/gartner-names-microsoft-a-leader-in-2019-endpoint-protection-platforms-magic-quadrant/}} including many leading and visionary products, before and after executing the random and section-injection attacks.
In many cases, our section-injection attack is able to drastically decrease the detection rate (see, \eg, AV1, AV3, AV7 and AV9), significantly outperforming the random attack (see, \eg, AV1 and AV9).
The reason may be that some of these antivirus programs already use static machine learning-based detectors to implement a first line of defense when protecting end-point clients from malware, as also confirmed in their blog or website, and this makes them more vulnerable to our attacks.
To conclude, our analysis highlights that these commercial products can be evaded with a transfer attack, and we believe that their detection rate could decrease even more with an optimized attack against them.

\section{Related work}
\label{sec:related}

Previous work is significantly different from \algoname, as it considers different settings and solutions. In particular, related approaches explore the creation of adversarial examples for information-security detectors, leveraging both gradient-based and black-box algorithms, as detailed in the following.

\myparagraph{Reinforcement learning.} Anderson et al.~\cite{anderson2017evading} propose a reinforcement learning approach to decide the best sequence of manipulation that leads to evasion.
To test the effectiveness of the agent, they also test the application of manipulations picked at random.
The model they used as a baseline is a primordial version of the GBDT classifier we have analyzed in this work, trained on fewer samples.
To train the policy of the learning agent, they let the model explore the space of adversarial examples, by fixing a budget for the number of queries that can be used.
The mean number of queries applied for training these policies is roughly 1600~\cite{anderson2017evading}.
The authors do not report the resulting file size of the adversarial malware: the reinforcement learning method contains actions that enlarge the representation on disk, but it is not clear how and how much.
Differently, our approach does not need a training phase, as it can be deployed as-is against the remote detector.
The transformation we use are functionality-invariant by design, and their application do not alter the execution flow of the program.
Lastly, we take into account how much content is added to the input malware, by plugging a regularizer inside the optimization process.
In this way, the amount of inserted noise is controlled, and the algorithm can find adversarial examples that not only evade the target classifier, but also they are limited in size.

\myparagraph{Genetic strategies.} Castro et al.~\cite{castro2019armed, labaca-castro2019aimed} apply both a random and genetic algorithm to perturb the input malware, and they test the functionality of the samples at each iteration of the optimization process inside a sandbox.
The mutations are the same proposed by Anderson et al.~\cite{anderson2017evading}.
The authors of these work state that they need approximately almost 4 minutes for creating adversarial malware, using 100 queries.
No architecture details have been unveiled.
We do not need to validate the malware inside a sandbox, as we include domain knowledge inside the mutation process.
For this reason, our methods performs 1,400 queries in the same time span.
They also do not report which are the most influential mutations that lead to evasion: the latter is crucial, we are dealing with potential vulnerabilities that lies inside statistical algorithms, whose presence is less evident compared to other security breaches.

\myparagraph{Generative Adversarial Networks}. Hu et al.\cite{hu2017generating} develop a Generative Adversarial Network (GAN)~\cite{goodfellow2014generative} whose aim is to craft adversarial malware that bypass a target classifier. 
The network learns which API imports should be added to the original sample, but no real malware is crafted, as that is attack only operates inside the feature space.
In contrast, we create functioning malware, as real samples are generated each time.

A recap of the black-box attacks against Windows malware detectors can be found in Table \ref{table:recap_blackbox}, where we compare the techniques we mentioned above with our method.

\begin{table}
\centering
{ 
\begin{tabular}{lclcc}
\textbf{Detector}                   & \textbf{FP} & \textbf{NS} & \textbf{AO} & \textbf{ST} \\ \hline
MalGAN~\cite{hu2017generating}      &             &             & \cmark      &             \\ \hline
ARMED~\cite{castro2019armed}        &             &             &             &             \\ \hline
AIMED~\cite{labaca-castro2019aimed} &             &             & \cmark      &             \\ \hline
RL Agent~\cite{anderson2017evading} & \cmark      & \cmark      & \cmark      &             \\ \hline
\textbf{\algoname}                           & \cmark      & \cmark      & \cmark      & \cmark      \\ \hline
\end{tabular}
}
\caption{Black-box adversarial attacks on Windows malware detectors. \textbf{FP:} Functionality-preserving transformations; \textbf{NS:} no sandboxing required; \textbf{AO:} attack optimization; \textbf{ST:}  attack stealthiness (e.g., file size optimization).
}
\label{table:recap_blackbox}
\end{table}

\section{Limitations and Open Issues}
\label{sec:limitations}
We discuss here which aspects of our work can be improved in the near future, by highlighting its current limitations.

\myparagraph{Countermeasures.} This work aims to show that learning-based detectors can be vulnerable to well-crafted attacks, even if they manipulate only a small fraction of the input program. We have however not investigated any potential mitigation strategy against our attacks.
One first line of defense may be to consider robust features against our attacks, \eg, features that are not affected by changes performed either at the byte or section level. For example, graph-based representations extracted from static analysis like Abstract Syntax Trees (ASTs) may be used to this end. However, extracting these features is typically much more computationally demanding and, at least in principle, practical transformations that can alter these features may also be derived and added to our optimization framework. 
A second line of defense may be to improve robustness of the learning algorithm~\cite{biggio18}, \eg, via adversarial re-training or by developing specific detection mechanisms for our attacks. It would also be interesting to study how a classifier could be hardened by embedding domain knowledge inside the training pipeline, defining loss functions that are invariant to the application of adversarial manipulations. 
Another interesting line of defense may be related to analyze the sequence of consecutive queries received from the same source, to detect whether a black-box attack performing correlated queries is taking place. We believe that all these defense strategies, especially if exploited in a complementary manner, may provide an interesting research direction towards designing more robust malware detectors.

\myparagraph{Dynamic classifiers.} It is finally worth remarking that our approach is clearly not effective against systems that use features computed by dynamically executing the input program, since the manipulations we applied only focus on the structure of an executable without altering its execution.
This issue may however be overcome by exploiting behavioral manipulations, including \emph{binary rewriting techniques}~\cite{wenzl2019hack}, \ie, approaches that modify the assembly code of a binary by adding new functionalities. 
Considering techniques that may affect dynamic analysis thus constitutes another interesting avenue to extend the impact of our work in the near future.

\section{Conclusions and Future work}
\label{sec:conclusion}

In this paper, we have presented a novel family of black-box attacks on learning-based Windows malware detectors that are both query-efficient and functionality-preserving, overcoming the limitations of previous work. Our attacks rely on the injection of benign content (which will never be executed) either at the end of the malicious file, or within newly-created sections, exploiting the ambiguities of the file format used to store programs on disk, without altering its execution traces.
The proposed attacks are formalized as a constrained minimization problem which enables optimizing the trade-off between the probability of evading detection and the size of the injected payload.  
Our extensive empirical evaluation on two popular learning-based Windows malware detectors has shown that our black-box attacks can bypass them with only few queries and very small payloads, even when the target models only output the predicted labels. We have also shown that our attacks can successfully transfer to other commercial antivirus solutions, finding that they can evade, on average, up to 12 commercial antivirus engines available on VirusTotal.
Nevertheless, we believe that a optimizing our attacks directly against these detectors might be even more effective.

\myparagraph{Future Work.} An interesting avenue for future work is related to investigating the applicability of suitable countermeasures against our attacks, as those discussed in Sect.~\ref{sec:limitations}, including the use of more robust feature representations (insensitive to byte-based or section-based manipulations) and learning paradigms (via adversarial re-training, specific attack detection mechanisms or the use of domain-knowledge constraints). 
Another promising research direction is to extend our attack beyond manipulations that only inject content either at the end of the file or within some newly-created sections. We firmly believe that this can be readily achieved, as our approach is already general enough to encompass a wider range of functionality-preserving manipulations, including those discussed in Sect.~\ref{subsec:realizable-attacks}.
Extending our work to deal with manipulations that can also modify the dynamic execution of malware programs, such as altering their control flow while preserving their malicious intent, is definitely challenging. Nevertheless, this would certainly provide an important step towards improving both the evaluation and the adversarial robustness of malware detectors trained on features extracted from dynamic program analysis.

%

\section*{Acknowledgement}
This work was partly supported by the PRIN 2017 project RexLearn (grant no. 2017TWNMH2), funded by the Italian Ministry of Education, University and Research.


%





\ifCLASSOPTIONcaptionsoff
  \newpage
\fi






\newpage
\begin{IEEEbiography}[{\includegraphics[width=1in,height=1.25in,clip,keepaspectratio]{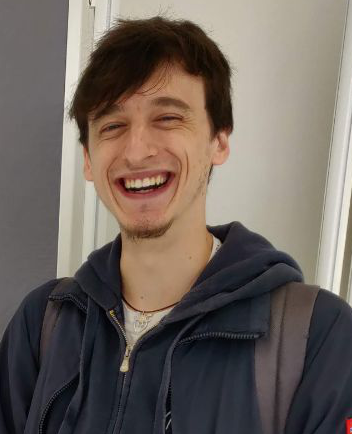}}]{Luca Demetrio}
received the M.Sc. degree (Hons.) and the Ph.D. degree in Computer Science from the University of Genoa, Italy, in 2017 and 2021.
His research interests cover the area of adversarial machine learning, with strong focus on its application in the cyber-security domain.
He is currently studying the weaknesses of threat detectors implemented with machine learning techniques, and how to exploit such vulnerabilities.
\end{IEEEbiography}
\vspace{-1cm}
\begin{IEEEbiography}[{\includegraphics[width=1in,height=1.25in,clip,keepaspectratio]{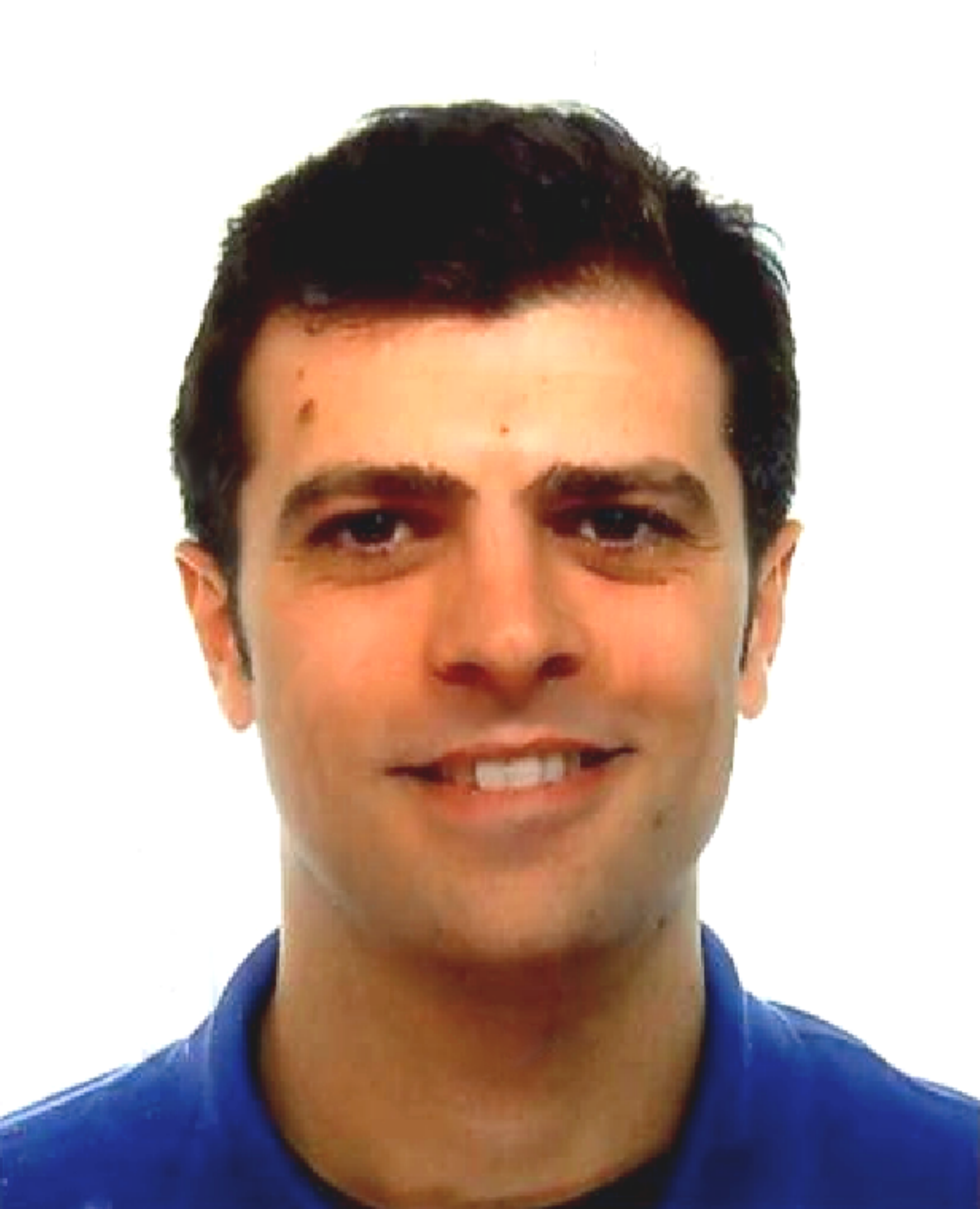}}]{Battista Biggio}
received the M.Sc. degree (Hons.) in Electronic Engineering and the Ph.D. degree in Electronic Engineering and Computer Science from the University of Cagliari, Italy, in 2006 and 2010. 
Since 2007, he has been with the Department of Electrical and Electronic Engineering, University of Cagliari, where he is currently an Assistant Professor. 
In 2011, he visited the University of Tuebingen, Germany, and worked on the security of machine learning to training data poisoning. 
His research interests include secure machine learning, multiple classifier systems, kernel methods, biometrics and computer security.
Dr. Biggio serves as a reviewer for several international conferences and journals. He is a senior member of the IEEE and a member of the IAPR.
\end{IEEEbiography}
\vspace{-1cm}
\begin{IEEEbiography}[{\includegraphics[width=1in,height=1.25in,keepaspectratio]{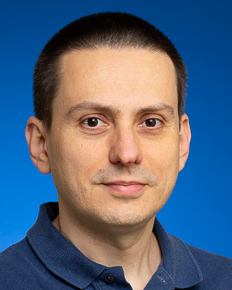}}]{Giovanni Lagorio}
received his M.Sc. (Hons) degree and Ph.D. in Computer Science from the University of Genoa, Italy, in 1999 and 2004. In 2000 he joined DIBRIS, where he started working on the design of object-oriented languages and systems.
He is currently an Assistant Professor at the University of Genoa, and his research interests have shifted more towards the security aspects of programs and systems; in particular, his interests include binary analysis and exploitation, adversarial machine learning, and ethical hacking.
\end{IEEEbiography}
\vspace{-1cm}
\begin{IEEEbiography}[{\includegraphics[width=1in,height=1.25in,clip,keepaspectratio]{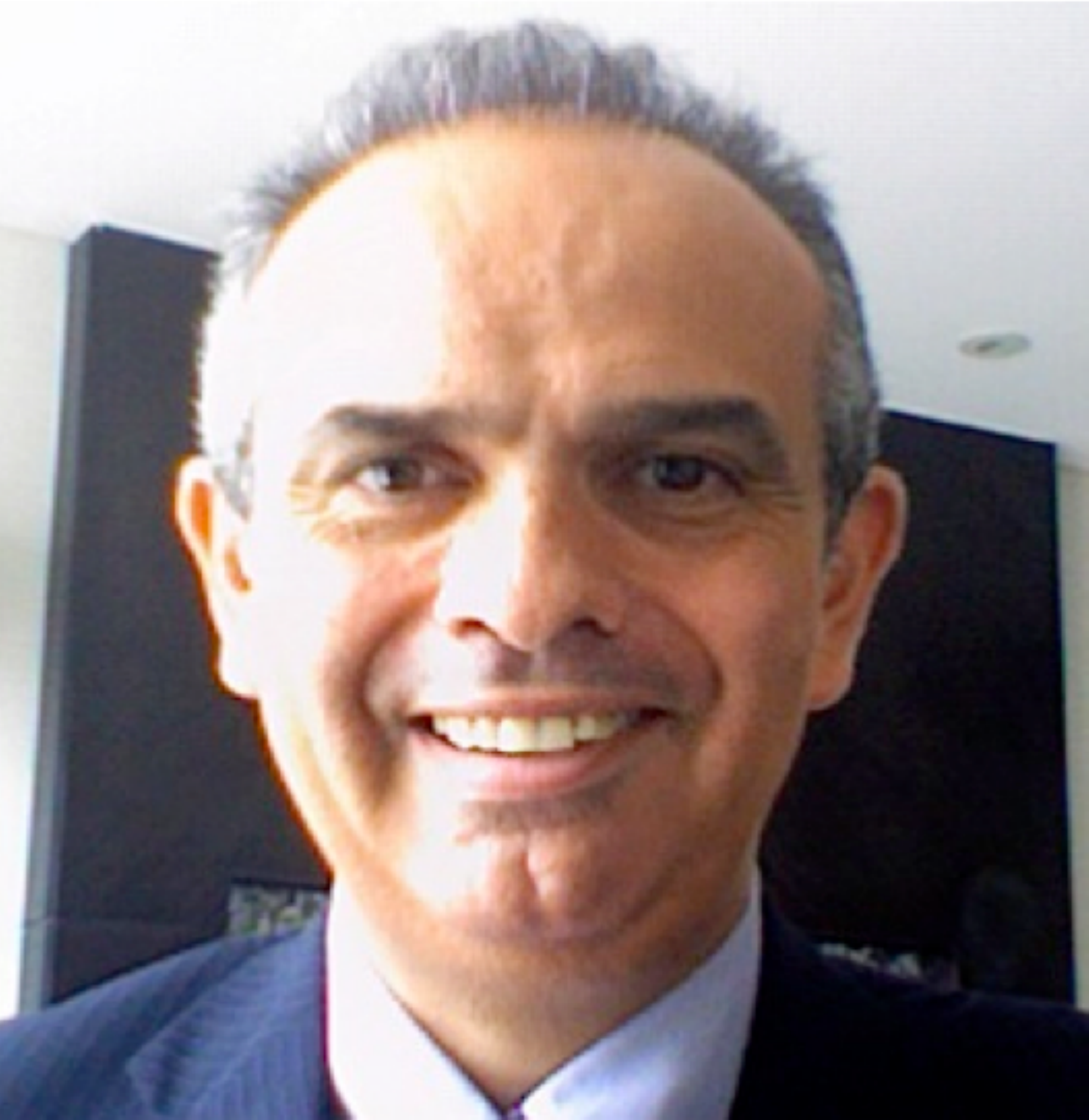}}]{Fabio Roli}
received his Ph.D. in Electronic Engineering from the University of Genoa, Italy. 
He was a research group member of the University of Genoa ('88-'94), and adjunct professor at the University of Trento ('93-'94). 
In 1995, he joined the Department of Electrical and Electronic Engineering of the University of Cagliari, where he is now professor of Computer Engineering and head of the research laboratory on pattern recognition and applications. 
His research activity is focused on the design of pattern recognition systems and their applications. 
He was a very active organizer of international conferences and workshops, and established the popular workshop series on multiple classifier systems. Dr. Roli is Fellow of the IEEE and of the IAPR.
\end{IEEEbiography}
\vspace{-1cm}
\begin{IEEEbiography}[{\includegraphics[width=1in,height=1.25in,clip,keepaspectratio]{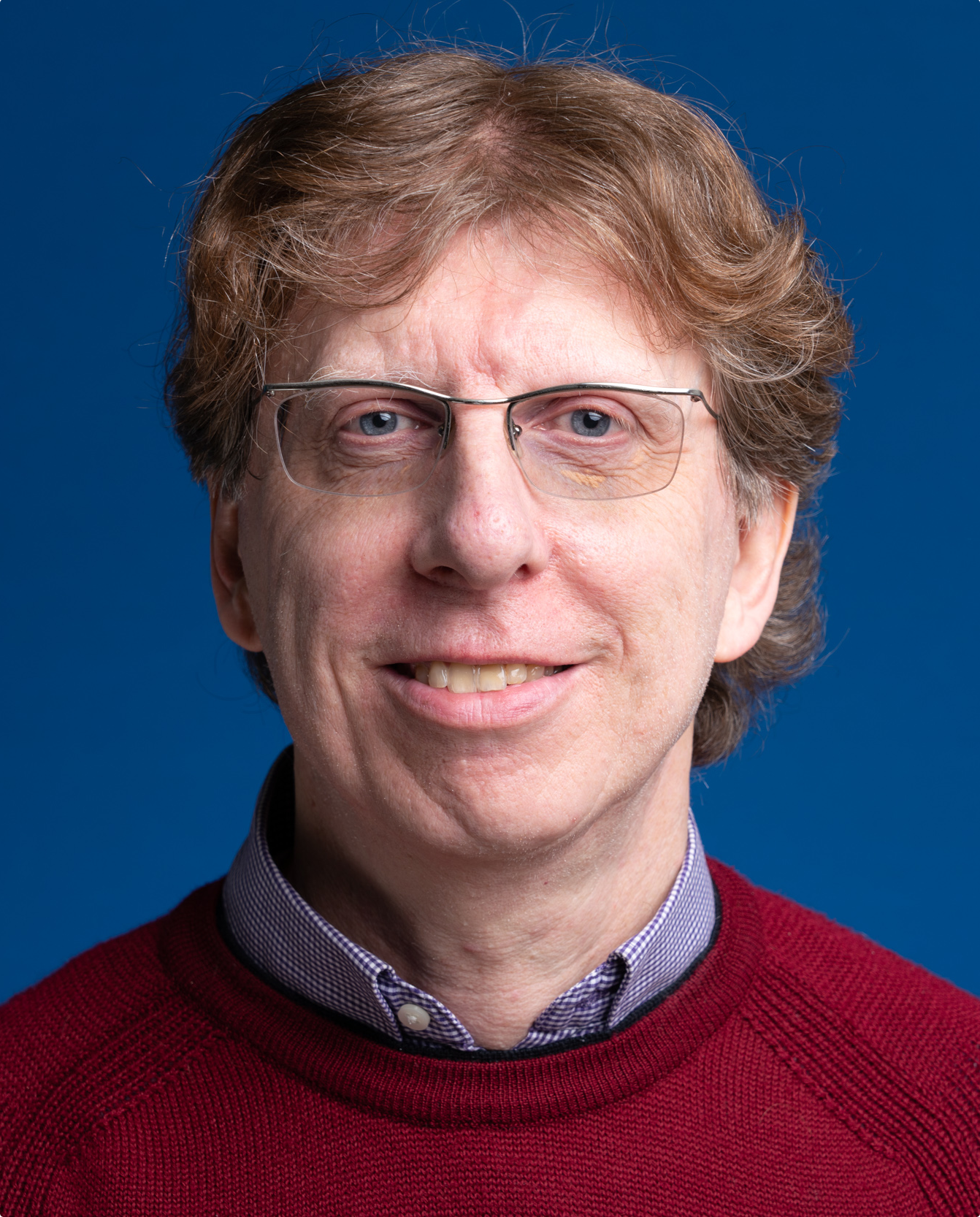}}]{Alessandro Armando}
received his PhD in Computer Engineering at the University of Genova.
His appointments include a position as research fellow at the University of Edinburgh and one at INRIA-Lorraine (France). 
He is professor at the University of Genova where he teaches Computer Security and has founded and coordinated the Master in Cybersecurity and Data Protection. 
In 2011 he founded (and led until 2016) the Security \& Trust Research Unit of the Bruno Kessler Foundation in Trento. 
He has been coordinator and team leader in several national and EU research projects, including the AVISPA, AVANTSSAR, SpaCIoS and SECENTIS projects. 
He contributed to the discovery of an authentication flaw in the SAML 2.0 Web-browser SSO Profile and of a serious man-in-the-middle attack on the SAML-based SSO for Google Apps. 
He is currently serving as vice director of the CINI National Cybersecurity Laboratory.
\end{IEEEbiography}

\balance

\end{document}